# Analyzing the Reduced Required BS Density due to CoMP in Cellular Networks

S. Alireza Banani, Raviraj S. Adve

Dept. of Electrical and Computer Engineering, University of Toronto, Toronto, ON, Canada
emails: alireza.banani@utoronto.ca, rsadve@comm.utoronto.ca

*Abstract*—In this paper we investigate the benefit of base station (BS) cooperation in the uplink of coordinated multi-point (CoMP) networks. Our figure of merit is the required BS density required to meet a chosen rate coverage. Our model assumes a 2-D network of BSs on a regular hexagonal lattice in which path loss, lognormal shadowing and Rayleigh fading affect the signal received from users. Accurate closed-form expressions are first presented for the sum-rate coverage probability and ergodic sum-rate at each point of the cooperation region. Then, for a chosen quality of user rate, the required density of BS is derived based on the minimum value of rate coverage probability in the cooperation region. The approach guarantees that the achievable rate in the entire coverage region is above a target rate with chosen probability. The formulation allows comparison between different orders of BS cooperation, quantifying the reduced required BS density from higher orders of cooperation.

## I. INTRODUCTION

With the aim of suppressing inter-cell interference (ICI) and improving the overall quality for the user, particularly at the cell edge, coordinated multi-point (CoMP) [1]-[2] techniques envision dynamic coordination of transmission/reception over a variety of different base stations (BSs) in frequency reuse-1 networks. CoMP techniques are categorized into joint processing (JP) and coordinated scheduling/beamforming (CS/CB) [3]-[4]. JP-CoMP leads to a larger capacity improvement over the CS/CB, since data is shared among multiple points for joint transmission. Many works have investigated the performance of JP-CoMP under different scenarios. These are too numerous to fully list here, but some representative examples are [5]-[14]. Of these examples, the behavior of CoMP under different precoding methods and perfect channel knowledge is investigated in [5]-[7]. The works in [8]-[11] studied the effect of imperfect channel state information (CSI) and limited feedback on the capacity of CoMP networks. The performance of CoMP schemes in heterogeneous networks have also been evaluated in [12]-[14]. However, no tractable formulations are provided in these networks and the results are largely limited to simulations.

Despite the large body of CoMP literature, there are modeling aspects which justify further investigation since they can lead to new insights into system design. For example, no one has, as yet, formally formulated the CoMP capacity while considering the simultaneous effects of path loss, shadowing and small-scale fading in the signal fluctuations. In this paper, we account for all these fading factors in a JP-CoMP multiuser uplink network with the specific aim of analyzing how CoMP allows for the use of fewer BSs in any geographical area while maintaining a chosen quality of service metric.

Here we provide analytical expressions for the sum-rate coverage probability (RCP) and ergodic sum-rate, for the case of a hexagonal grid-based network. The rate expressions are presented in the general format for the cooperation amongst $N$ BSs and depend heavily on users' location in the cooperation region. For typical values of cooperation order $N \leq 3$, the minimum user rate corresponds to the case when all the users are positioned at the locations experiencing equal path loss from the BSs. The minimum user rate in the entire cooperation region is useful for designing a CoMP network subject to a quality of rate for each user. In this paper, the worst user rate is used as the basis for finding the required BS density for maintaining the RCP (or achievable rate) at all points of the network above a target value.

This paper extends the work in [15] where cooperation amongst only two BSs is considered. General expressions are provided for cooperation among $N$ BSs. Thus, the formulation allows comparison between the results of different orders of BS cooperation, including the no-cooperation case (no CoMP). In particular, we quantify the gains in the reduced required BS density from increasing orders of cooperation. For brevity, perfect CSI is assumed to be available at the BSs. Also, we assume a high-speed backhaul for information exchange (data, control, synchronization, and CSI) between the BSs. The rest of the paper is organized as follows. Section II describes the JP-CoMP uplink system model. The analytical rate expressions are provided in section III, followed by supporting simulations in section IV. Finally the conclusion is drawn in Section V. The notation is conventional, as follows. Matrices are in bold capitals and vectors in bold lower case. The notations $(\cdot)^H$, and $(\cdot)^T$ denote the conjugate transpose, and transpose, respectively. A vector $\mathbf{a} \sim CN(0,1)$ contains i.i.d. zero-mean complex Gaussians with variance 1.

## II. SYSTEM MODEL

We consider a large, cellular, CoMP uplink network where BSs are arranged according to a hexagonal grid. $N$ adjacent BSs constitute a cooperation region and serve all the users which fall in that cooperation region. Figs. 1-(a) and 1-(b) show sample cooperation regions for adjacent BSs cooperating with $BS_1$ in a network with cooperation order $N = 2$, and $N = 3$, respectively. We assume use of resource allocation scheme that eliminates interference between cooperation regions; e.g., single carrier frequency domain multiple access (SC-FDMA), can be used to assign orthogonal frequencies.

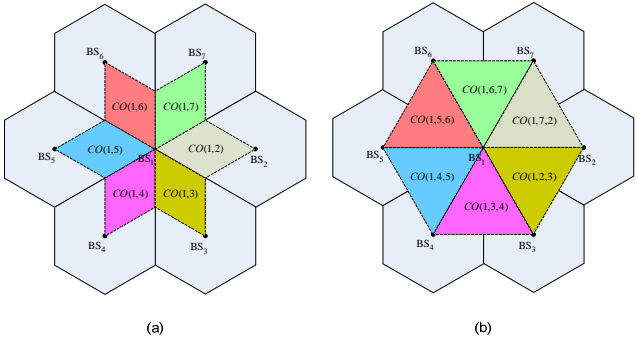

Figure 1. The cooperation regions associated with $BS_1$ for: (a) cooperation order $N = 2$; (b) cooperation order $N = 3$: $CO(1,j,k)$ denotes the cooperation region between $BS_1$, $BS_j$, and $BS_k$.

The cooperation of $N$ BSs (each with $M$ receive antennas) processes data from $U$ users all within its cooperation region, and forms a virtual $NM \times U$ multiple-input, multiple-output (MIMO) frequency selective fading channel. For brevity, it is assumed that each user is equipped with a single transmit antenna. This is not restrictive and the formulation can be modified for users with more than one antenna. With the use of SC-FDMA, it is sufficient to observe the baseband transmission on a single frequency-flat subcarrier. Under perfect synchronization, the transmission and equalization of symbols on an SC-FDMA subcarrier can be expressed as

$$\hat{\mathbf{s}} = \mathbf{W}\mathbf{y} = \mathbf{W}(\mathbf{Hs}+\mathbf{n}) = \underbrace{\mathbf{WHs}}_{signal} + \underbrace{\mathbf{Wn}}_{interference + noise}, \quad (1)$$

where $\mathbf{s} = [s_1 \cdots s_U]^T$ contains the symbols transmitted by the users. As we do not consider user cooperation in this paper, the signals originating from different users are uncorrelated. Furthermore, with no transmitter-side channel knowledge, users with equal transmit power are considered. Thus, the signal covariance matrix $E\{\mathbf{ss}^H\} = \Phi_{\mathbf{ss}} = \sigma_s^2 \mathbf{I}$ where $\sigma_s^2$ is subject to a user power constraint. The $2M \times U$ matrix $\mathbf{H}$ is a realization of the channels on this subcarrier with the elements modeled as

$$[\mathbf{H}]_{i,u \atop u=1,\cdots,U} = \frac{\bar{h}_{i,u}}{10^{(\text{PL}(d_n^u)+L_n^u)/20}} \quad ; \quad \begin{array}{l} i = (n-1)M+1,\cdots,nM \\ n = 1,\cdots,N \end{array}. \quad (2)$$

where $\bar{h}_{i,u} \sim CN(0,1)$ represents the normalized complex channel gain reflecting small-scale Rayleigh fading. With uncorrelated receive antennas at each BS, $\bar{h}_{i,u}$ are independent for all $i$ and $u$. $\text{PL}(d_n^u)$ is the path loss (in dB) from the $u$-th user to the $n$-th cooperative BS, located at distance $d_n^u$, and can be expressed as $\text{PL}(d_n^u) = a + b\log_{10} d_n^u$. Here, $a$ and $b$ are constant coefficients chosen based on the path loss model used. The effect of shadowing or large-scale fading is reflected in $L_n^u \sim CN(0,\sigma_L)$, representing the power fluctuation around the average path loss from $n$-th cooperative BS to the $u$-th user.

In (1), $\mathbf{y} = [y_1 \cdots y_{NM}]^T$ is the signal vector received by the $N$ cooperating BSs, and includes the effect of zero-mean Gaussian noise $\mathbf{n} \sim CN(0,\sigma^2)$ with covariance $\Phi_{\mathbf{nn}} = \sigma^2 \mathbf{I}$; this data is then equalized, via the matrix $\mathbf{W}$, for data recovery. The structure of $\mathbf{W}$ depends on the particular CoMP strategy. Here, we use the receive zero forcing (ZF) filter because of its pragmatic tradeoff between simplicity and performance. The receive ZF filter combines matched filtering and interference suppression [16] and can be expressed as $\mathbf{W} = (\mathbf{H}^H \mathbf{H})^{-1} \mathbf{H}^H$. With this structure, it is inherently assumed that the number of users on each subcarrier associated with each cooperation region satisfies $U \leq NM$.

### III. ACHIEVABLE RATE

The achievable sum-rate (in b/s/Hz) within each cooperation region with $U$ users is given as,

$$R = \sum_{u=1}^{U} R_u = \log_2 \left| \mathbf{I}_U + \mathbf{R}_{int+noise}^{-1} \mathbf{R}_{signal} \right| = \log_2 \left| \mathbf{I}_U + \frac{\sigma_s^2}{\sigma^2} \mathbf{H}^H \mathbf{H} \right|, \quad (3)$$

with

$$\mathbf{R}_{signal} = \sigma_s^2 \mathbf{WHH}^H \mathbf{W}^H = \sigma_s^2, \quad (4)$$

$$\mathbf{R}_{int+noise} = \sigma^2 \mathbf{WW}^H = \sigma^2 (\mathbf{H}^H \mathbf{H})^{-1}. \quad (5)$$

The sum-rate formula in (3) is lower-bounded by $R' = \log_2 |(\sigma_s^2/\sigma^2)\mathbf{H}^H\mathbf{H}|$. On the other hand, $R'$ itself can be upper-bounded by $R'' = \log_2 \prod_{u=1}^{U} (\sum_{i=1}^{NM} (\sigma_s^2/\sigma^2) \| [\mathbf{H}]_{i,u} \|^2)$. While, in general, there is no guarantee that $R'' \leq R$ for any location within the cooperation region, simulations suggest that when users are far from BSs (the received signal SINR from each user is low), $R'' \leq R$ is achieved. Therefore, sum-rate $R$ can be approximated by $R''$. Correspondingly, the sum RCP which is defined as the probability that users can achieve sum-rate above a threshold, is approximated as

$$P(R > T) \cong P(R'' > T)$$

$$= P(\log_2 \prod_{u=1}^{U} \left(\sum_{i=1}^{NM} (\sigma_s^2/\sigma^2) \| [\mathbf{H}]_{i,u} \|^2 \right) > T), \quad (6)$$

$$= P(\prod_{u=1}^{U} \underbrace{\left(\sum_{i=1}^{NM} (\sigma_s^2/\sigma^2) \| [\mathbf{H}]_{i,u} \|^2 \right)}_{SNR(u)} > 2^T)$$

where

$$SNR(u) = \sum_{n=1}^{N} \frac{\sigma_s^2}{\sigma^2} 10^{-a/10} \left( \sum_{i=(n-1)M+1}^{nM} \| h_{i,u} \|^2 \right) (d_n^u)^{-b/10} 10^{-L_n^u/10}, \quad (7)$$

is the instantaneous signal-to-noise-ratio (SNR) associated with user $u$ under maximum-ratio-combining. $SNR(u)$ depends on the location of user via $\text{PL}(d_n^u)$ as well as the instantaneous realizations of $\bar{h}_{i,u}$; $i = (n-1)M+1,\cdots,nM$, and $L_n^u$. By changing the variables in (7), the $SNR(u)$ can be written as

$$SNR(u) = \sum_{n=1}^{N} \underbrace{c \omega_n z_n}_{\xi_n}, \quad (8)$$

with $c = (\sigma_s^2/2\sigma^2) \times 10^{-a/10}$, $\omega_n = \sum_{i=(n-1)M+1}^{nM} \|h_{i,u}\|^2$, $\alpha = b/10$, and $z_n = (d_n^u)^{-\alpha} 10^{-L_n^u/10}$. Here, $z_n$ is a log-normal random variable with $z_n \sim LN(\mu_{z_n} = -\alpha \ln d_n^u, \sigma_{z_n} = (0.1 \ln 10)\sigma_L)$. The random variable $\omega_n$ has a chi-squared distribution with $2M$ degrees of freedom ($\omega_n \sim \chi^2(2M)$) since it is the sum of the squares of $2M$ independent standard normal distributions with unit variance. Finally, from $\xi_n = c\omega_n$, it follows that $\xi_n \sim \Gamma(\kappa, \theta)$ has a gamma distribution with parameters $\kappa = 2M/2$ and $\theta = 2c = (\sigma_s^2/\sigma^2) \times 10^{-a/10}$.

From (8), $SNR(u)$ is, in fact, a linear combination of the log-normal random variables $z_n; n = 1, \cdots, N$ with the coefficients $\xi_n; n = 1, \cdots, N$ which, themselves, are independent random variables. The key to simplifying this expression is to realize that linear combinations of log-normal random variables can be close approximated as a single log-normal random variable [17]. The work in [17] presents several such approximations based on a generalization of the Moment Matching. For example, by matching the first and second moments, the log-normal approximation of $SNR(u)$ is

$$SNR(u) \sim LN(\mu_{SNR(u)}, \sigma_{SNR(u)}) \quad (9)$$

with $\mu_{SNR} = 2\ln(\beta_1) - 0.5\ln(\beta_2)$ and $\sigma_{SNR}^2 = -2\ln(\beta_1) + \ln(\beta_2)$, where

$$\beta_1 = \sum_{n=1}^{N} E\{\xi_n\} \exp(\mu_{z_n} + \sigma_{z_n}^2/2) \quad (10)$$

$$\beta_2 = \sum_{n=1}^{N} E\{\xi_n^2\} \exp(2\mu_{z_n} + 2\sigma_{z_n}^2) + \sum_{i=1}^{N}\sum_{\substack{j=1 \\ j \neq i}}^{N} E\{\xi_i\}E\{\xi_j\}\exp(\mu_{z_i} + \mu_{z_j} + (\sigma_{z_i}^2 + \sigma_{z_j}^2)/2) \quad (11)$$

The gamma distributed random variables $\xi_n$ have a mean and second order moment of $E\{\xi_n\} = \kappa\theta$ and $E\{\xi_n^2\} = \kappa\theta^2 + \kappa^2\theta^2$, respectively. Therefore,

$$\beta_1 = M \frac{\sigma_s^2}{\sigma^2} 10^{-a/10} e^{(\sigma_z^2/2)} \sum_{n=1}^{N} (d_n^u)^{-\alpha}, \quad (12)$$

$$\beta_2 = M \frac{\sigma_s^4}{\sigma^4} 10^{-2a/10} e^{\sigma_z^2} \left[ (M+1)e^{\sigma_z^2} \sum_{n=1}^{N}(d_n^u)^{-2\alpha} + M \sum_{i=1}^{N}\sum_{j=1, j\neq i}^{N} (d_i^u)^{-\alpha}(d_j^u)^{-2\alpha} \right]. \quad (13)$$

The product of all $SNR(u); u = 1, \cdots, U$ is a log-normal distribution, i.e., $\prod_{u=1}^{U} SNR(u) \sim LN(\sum_u \mu_{SNR(u)}, (\sum_u \sigma_{SNR(u)}^2)^{1/2})$. As a result, the sum RCP in (6) can be approximated as follows:

$$P(R > T) \cong P(\prod_u SNR(u) > 2^T) \approx Q\left(\frac{T\ln 2 - \sum_u \mu_{SNR(u)}}{(\sum_u \sigma_{SNR(u)}^2)^{1/2}}\right). \quad (14)$$

The ergodic sum-rate can also be achieved from the sum-rate coverage probability. Since for a positive random variable $X$, $E\{X\} = \int_{t>0} P(X > t)dt$, an approximation of ergodic sum-rate is given as

$$R_{ergodic} = \int_{T>0} P(R > T) dT \cong \int_{T>0} Q\left(\frac{T\ln 2 - \sum_u \mu_{SNR(u)}}{(\sum_u \sigma_{SNR(u)}^2)^{1/2}}\right) dT. \quad (15)$$

The above integration can be solved with the use of an accurate approximation of $Q(x)$ given by [18],

$$Q(x) \approx e^{-x^2/2} \sum_{j=1}^{J} a_j x^{j-1} \; ; \; x \geq 0, \quad (16)$$

where 

$$a_j = \frac{(-1)^{j+1} \overline{A}^j}{\overline{B}\sqrt{\pi}(\sqrt{2})^{j+1} j!}. \quad (17)$$

with $\overline{A} = 1.98$ and $\overline{B} = 1.135$ (an accurate approximation is obtained for $J = 10$ at all values of $x \geq 0$). Let $\overline{a} = \sum \mu_{SNR(u)}$, and $\overline{b} = (\sum \sigma_{SNR(u)}^2)^{1/2}$. We get

$$R_{ergodic} \cong \int_{T>0} Q\left(\frac{T\ln 2 - \overline{a}}{\overline{b}}\right) dT$$
$$= \int_{T=0}^{\overline{a}/\ln 2} \left[1 - Q\left(\frac{\overline{a} - T\ln 2}{\overline{b}}\right)\right] dT + \int_{\overline{a}/\ln 2}^{\infty} Q\left(\frac{T\ln 2 - \overline{a}}{\overline{b}}\right) dT$$
$$= \frac{\overline{a}}{\ln 2} - \int_{T=0}^{\overline{a}/\ln 2} e^{-\left(\frac{\overline{a}-T\ln 2}{\overline{b}}\right)^2/2} \sum_j a_j \left(\frac{\overline{a} - T\ln 2}{\overline{b}}\right)^{j-1} dT$$
$$+ \int_{\overline{a}/\ln 2}^{\infty} e^{-\left(\frac{T\ln 2 - \overline{a}}{\overline{b}}\right)^2/2} \sum_j a_j \left(\frac{T\ln 2 - \overline{a}}{\overline{b}}\right)^{j-1} dT \quad (18)$$

By setting $y \equiv ((T\ln 2 - \overline{a})/\overline{b})^2/2$,

$$R_{ergodic} \cong \frac{\overline{a}}{\ln 2} - \frac{\overline{b}}{\ln 2} \int_0^{(\overline{a}/\overline{b})^2/2} e^{-y} \sum_j a_j (2y)^{(1/2)j-1} dy$$
$$+ \frac{\overline{b}}{\ln 2} \int_{(\overline{a}/\overline{b})^2/2}^{\infty} e^{-y} \sum_j a_j (2y)^{(1/2)j-1} dy$$
$$= \overline{a}/\ln 2 - (\overline{b}/\ln 2)\sum_{j=1}^{J} 2^{(1/2)j-1} a_j \Gamma(0.5j)$$
$$+ (\overline{b}/\ln 2)\sum_{j=1}^{J} 2^{(1/2)j-1} a_j \Gamma(0.5j, (\overline{a}/\overline{b})^2/2), \quad (19)$$

where $\Gamma(x) = \int_0^{\infty} t^{x-1} e^{-t} dt$, and $\Gamma(x,y) = \int_y^{\infty} t^{x-1} e^{-t} dt$ are the standard gamma and incomplete gamma functions, respectively. We note that both expressions in (14) and (19) are point-wise in the sense that they provide approximations for the sum RCP and the ergodic sum-rate at each point of the coverage region, respectively. The minimum of the sum RCP and/or ergodic sum-rate have particular importance in the design of a network [19]-[20]. In this paper, the minimum RCP is used towards obtaining the required density of BSs with the objective of ensuring that the achievable rate at all points in the network is above a target rate with a chosen probability. For the example of two BS cooperation ($N = 2$) with uniform diamond-shaped cooperation regions in Fig. 1(a), it is shown in [15] that the minimum sum RCP is obtained at the points located at the two side edges with $d_1^u = d_2^u = \sqrt{3}D/3$ where $D$ is the distance between two BSs. In a similar manner it can be shown that for $N = 3$, the minimum sum RCP (or ergodic sum-rate) corresponds to the case when all the users are gathered at the centre of the triangular cooperation region experiencing the same channel and path loss from the three BSs ($d_1^u = d_2^u = d_3^u = \sqrt{3}D/3$). As a result, for $N \leq 3$, the worst RCP and worst ergodic rate associated with each user is obtained as

$$P(R_u^w > \underbrace{T/U}_{\tilde{T}}) = P(R^w > T), \quad (20)$$

$$R_{u,ergodic}^w = R_{ergodic}^w / U, \quad (21)$$

where the superscript '$w$' refers to the value at the worst-case point and $R_u^w$ is an approximation of the worst achievable rate of each user. It is worth noting that the choice of $T/U$ in (20) is conservative (assuming all $U$ users get the same minimum rate). However, this allows for a simple analysis; the expressions are general enough that one could choose any rate value in the outage expression in (14).

## IV. SIMULATION RESULTS

We note that the worst user RCP and the ergodic rate in (20)-(21) are functions of $D$ or, the BS density $\lambda = 1/\text{cell Area} = 1/(\sqrt{3}D^2/2)$. This result makes it possible for the designer to adjust the density of BSs in the network based on a target value for RCP and/or ergodic rate. Here we illustrate how the analysis in Section III can be used to choose a required BS density in CoMP networks.

The presented design approach is simulated for the example of $N \leq 3$. The algorithm is applicable for different orders of BS cooperation, number of users and BS antennas, but the examples provided here are sufficient to illustrate the approach. The details of the simulations are: the nominal carrier frequency is 2 GHz; The channel matrix **H** is generated according to (2) with the LTE-A outdoor micro path loss $PL(d_n^u) = -39 + 67\log_{10}(d_n^u)$ [2], and log-normal shadow fading with $\sigma_L = 6$ dB; the power of each user is set to $\sigma_s^2 = 30$ dBm; and the additive noise at each BS antenna is $\sigma^2 = -90$ dBm. We note that the parameter values in the simulations are just for the illustration purposes and any other values only scale the result.

Here, an example is presented to show how the required BS density is chosen for a target value of RCP. We consider an uplink CoMP network with $N = 3$ and three users per cooperation region. We require the user RCP to be greater than 0.7 for the rate of 1 b/s/Hz in the entire cooperation region, i.e., we require that every point in the cooperation region receive at least a rate of 1 b/s/Hz with probability 0.7. Figure 2 illustrates the associated worst RCP for different values of BS density. The results are depicted for different number of antennas at each BS, $M$. For a given $M$ and the desired target value of RCP, the optimal BS density can be read from this figure. For the chosen RCP value of 0.7 with $M = 1$, the required density is $\lambda = 7.5 \times 10^{-6}$ BS/m$^2$ (this value corresponds to $D = 390$ m). As expected, the required BS density decreases with an increase in $M$. In other words, higher RCP is achieved for larger values of $M$. Since the number of antennas at the users are kept fixed in this example, the improvement in RCP is the result of average increase in the SNR at the receiver (referred to as array gain) arising from coherent combining effect of multiple antennas at the BSs. However, the increase in the array gain diminishes for large values of $M$.

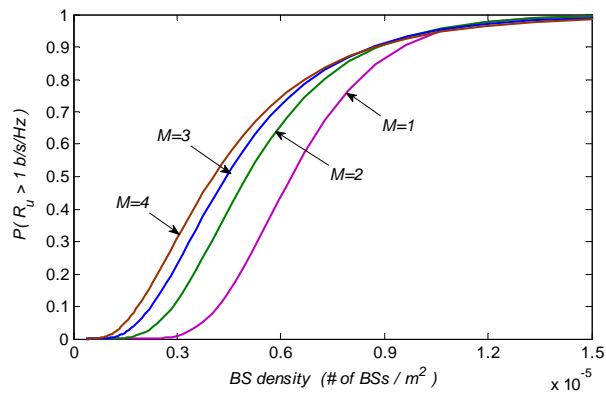

Figure 2. The worst RCP versus BS density for different numbers of antennas at each BS, $M$, in an uplink CoMP network with three BS cooperation and 3 users per cooperation region.

To illustrate the resulting RCP for the rate of 1 b/s/Hz in a 2-dimensional cooperation region with $\lambda = 7.5 \times 10^{-6}$ BS/m$^2$, and $M = 1$, users 2 and 3 are located at the centre of the triangular cooperation region (worst-case point) and user 1 is placed at different points of the cooperation region. Fig. 3 shows the RCP (for the rate of 1 b/s/Hz) of user 1 at different locations. The contour plot in the figure indicates the locations of user 1 with the same RCP. Fig. 3 shows that, as required by the analysis, the RCP is always higher than 0.7 and reaches its minimum when user 1 is located near the other 2 users at the centre. We note that, since the design is based on $R''$ (which lower-bounds $R$ when users are far from BSs) rather $R$, a higher RCP of 0.75 is achieved at the worst-case point.

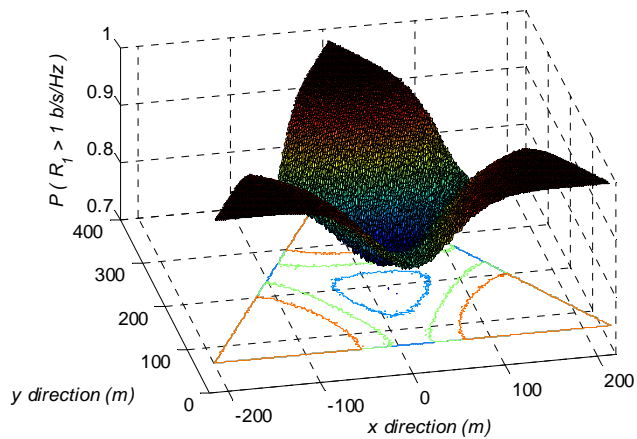

Figure 3. RCP (for the rate of 1 b/s/Hz) associated with user 1 in a cooperation region of three BSs located at $(-D,0)$, $(+D,0)$, and $(0, \sqrt{3}D/2)$ with $D = 390$ m, $M = 1$, and $U = 3$; Users 2 and 3 are located at the centre and user 1 is placed at different points of the cooperation region.

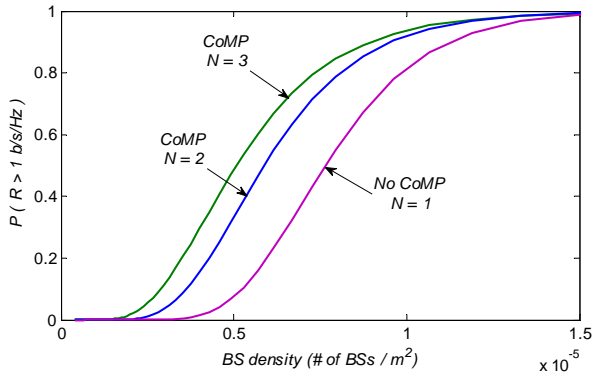

Figure 4. Comparison of the required BS density in the uplink networks with different orders of BS cooperation for different values of RCP (for the rate of 1 b/s/Hz). There are one antenna per BS and one user per cooperation region.

Next, comparison is made between the results of networks with different order of BS cooperation. This include the no-cooperative case (no CoMP) in which users within each cell is governed by its own BS, only. We note that each network has to satisfy $U \leq NM$. As a result, we consider the simplest case of $M = 1$, with one user per cooperation region. The associated RCP for the rate of 1 b/s/Hz, is illustrated in Fig. 4 for different values of BS density for $N \leq 3$. As expected, the required BS density decreases with the cooperation order $N$ for a given RCP. However, the greatest gain in the reduced required BS density, is obtained by going from no cooperation ($N$=1) to CoMP with $N = 2$ and smaller returns are obtained for going from $N = 2$ to $N = 3$ (with diminishing returns for large values of $N$, not shown here).

As an added benefit, the presented formulation allows us to quantify this gain for a given target value of RCP. In particular, for the example of RCP = 0.7, the network with $N = 2$ ($N = 3$) needs a factor of 0.82 (0.69) the required BS density in a no-cooperative network.

## V. SUMMARY AND CONCLUSIONS

In this paper, at first, we obtain analytical expressions for the sum-RCP and the ergodic sum-rate in uplink CoMP hexagonal-grid-based networks with any orders of BS cooperation. Our model considers all the important components of fading such as path loss, shadowing and small-scale fading. Then, the worst of user RCP is used as the figure of merit to finding the required BS density, for maintaining the RCP in entire cooperation region above a target value. As an added benefit, the formulations allow comparison between networks with different orders of BS cooperation, quantifying the gain in the reduced required BS density from higher orders of BS cooperation. The expressions are general enough to accommodate other figures of merit.

ACKNOWLEDGMENT

The authors would like to acknowledge the financial support of TELUS.